\documentclass{aa}  

\usepackage{graphicx}
\usepackage[varg]{txfonts}
\usepackage{units}
\usepackage{mathtools}
\usepackage{color}
\usepackage{natbib,twoopt}
\usepackage{hyperref} 
\bibpunct{(}{)}{;}{a}{}{,}             

\makeatletter
  \newcommandtwoopt{\citeads}[3][][]{\href{http://adsabs.harvard.edu/abs/#3}%
    {\def\hyper@linkstart##1##2{}%
     \let\hyper@linkend\@empty\citealp[#1][#2]{#3}}}
  \newcommandtwoopt{\citepads}[3][][]{\href{http://adsabs.harvard.edu/abs/#3}%
    {\def\hyper@linkstart##1##2{}%
     \let\hyper@linkend\@empty\citep[#1][#2]{#3}}}
  \newcommandtwoopt{\citetads}[3][][]{\href{http://adsabs.harvard.edu/abs/#3}%
    {\def\hyper@linkstart##1##2{}%
     \let\hyper@linkend\@empty\citet[#1][#2]{#3}}}
  \newcommandtwoopt{\citeyearads}[3][][]%
    {\href{http://adsabs.harvard.edu/abs/#3}
    {\def\hyper@linkstart##1##2{}%
     \let\hyper@linkend\@empty\citeyear[#1][#2]{#3}}}
  \renewcommand*\aa@pageof{, page \thepage{} of \pageref*{LastPage}}
\makeatother
\hypersetup{pdftitle = {Solid accretion onto planetary cores in radiative disks},
            pdfcreator = {\LaTeX},
            pdfauthor = {Apostolos Zormpas, Giovanni Picogna, Barbara Ercolano, Wilhelm Kley},
            pdfsubject = {},
            pdfview = {FitH},
            pdfstartview = {FitH},
            colorlinks = {true},
            linkcolor = [rgb]{0,0.35,0.7},
            citecolor = [rgb]{0,0.35,0.7},
            filecolor = [rgb]{0.61,0,0},
            urlcolor = [rgb]{0,0.35,0.7},
           }

\usepackage{afterpage}
\usepackage{silence}
\usepackage{siunitx}
\DeclareSIUnit\au{au}
\DeclareSIUnit\Msun{M$_\odot$}
\DeclareSIUnit\yr{yr}
\DeclareSIUnit\pp{pp}

\usepackage{bm}

\begin{document}


\title{Solid accretion onto planetary cores in radiative disks}
\subtitle{}

\author{
    Apostolos Zormpas\inst{1}
    \and
    Giovanni Picogna\inst{1}
    \and
    Barbara Ercolano\inst{1,2}
    \and
    Wilhelm Kley\inst{3}
}
\institute{
    Universit\"{a}ts-Sternwarte, Ludwig-Maximilians-Universit\"{a}t M\"{u}nchen, Scheinerstr.~1, 81679 M\"{u}nchen, Germany \\
    \email{zormpas@usm.lmu.de}
    \and
    Excellence Cluster Origin and Structure of the Universe, Boltzmannstr. 2, 85748 Garching bei M\"unchen, Germany
    \and
    Institut f\"{u}r Astronomie und Astrophysik, Universit\"{a}t T\"{u}bingen, Auf der Morgenstelle 10, 72076 T\"{u}bingen, Germany
}

\date{Last updated \today; in original form \today}

\abstract{The solid accretion rate, which is necessary to grow gas giant planetary cores within the disk lifetime, has been a major constraint for theories of planet formation. We tested the solid accretion rate efficiency on planetary cores of different masses embedded in their birth disk by means of 3D radiation-hydrodynamics, where we followed the evolution of a swarm of embedded solids of different sizes. We found that by using a realistic equation of state and radiative cooling, the disk at $5$ au is able to efficiently cool and reduce its aspect ratio. As a result, the pebble isolation mass is reached before the core grows to \SI{10}{M_\oplus}, thus fully stopping the pebble flux and creating a transition disk.
Moreover, the reduced isolation mass halts the solid accretion before the core reaches the critical mass, leading to a barrier to giant planet formation, and this explains the large abundance of super-Earth planets in the observed population.}



\maketitle

\titlerunning{Solid accretion onto planetary cores in radiative disks}
\authorrunning{Zormpas et al.}


\section{Introduction}

The formation of the solid core of gas giant planets is still an open question in regards to the core accretion model \citep[][]{pollack1996formation}.
The timescale on which a core of critical mass must be formed in order for runaway gas accretion to set in is given by the disk lifetime.
Depending on the stellar properties, this period has been found observationally to be between $3$ and \SI{10}{Myr} for Solar-type stars \citep[][]{ribas2015}.
On this short timescale, interstellar medium $\mu m$ dust grains must grow by almost $14$ orders of magnitude in size and $40$ orders of magnitude in mass.
The critical core mass value at which the hydrostatic equilibrium breaks down is in the $5-20\, M_\oplus$ range \citep{mizuno1980,bethune2019}, depending on how the envelope is able to cool \citep{rafikov2006,piso2015}, and it is strongly dependent on the adopted opacity \citep{mordasini2014,ormel2014,brouwers2019}.
Furthermore, due to the interaction between the planetary core and its natal circumstellar disk, planets migrate and can be accreted onto the central star if no other physical processes stop them.

In order to match the observed population of gas giants, a faster growth timescale is needed in the core accretion paradigm. In recent years, two promising mechanisms have been studied.
Under certain circumstances, the streaming instability might allow for the growth of planetesimal-sized objects directly from small dust particles \citep[see e.g.,][for a review]{johansen2014SIreview}, while pebble accretion might enhance the solid accretion on growing planetary cores \citep[][]{ormel2010,lambrechts2012}.

Although these mechanisms have successfully reduced the time required to form giant planetary cores, a further barrier must be overcome in order to reach the core mass necessary to ignite the rapid gas accretion phase.
A growing planet is able to open a partial gap in the surrounding gaseous disk by generating a pressure maximum, which effectively halts the inward drift of solids in the disk, in particular the pebble component.
The mass at which the pebble flux is halted has been defined as the pebble isolation mass \citep[][]{lambrechts2014separating}, and it varies based on the local physical properties of the disk \citep[][]{bitsch2018pebble,ataiee2018,picogna2018particle}.

Within this framework, we study the pebble accretion rate and isolation mass, in realistic 3D radiative hydrodynamical simulations, and how this affects the efficiency of gas giant planet formation.
In Sect.~\ref{sec:numerics} we describe the numerical set-up adopted to model the solid evolution and accretion onto a planetary core embedded in a disk with a realistic equation of state. We then discuss our results in Sect.~\ref{sec:results} and draw the main conclusions in Sect.~\ref{sec:conclusions}.

\section{Numerical methods}\label{sec:numerics}
We consider the same setup as in \cite{picogna2018particle} to allow for a direct comparison; the only exception is that instead of using a locally isothermal equation of state for the gas, we adopted the ideal equation of state and included the radiative transfer in the flux-limited diffusion approximation (FLD) \citep{kolb2013radiation}, and we consider laminar viscous disks.

\subsection{Gas component}\label{par:gasdisk}

The disk was initially set in an axisymmetric configuration ranging from $2.08$ to \SI{13}{au} ($0.4$-$2.5$ in code units, where the unit of length is \SI{5.2}{au}).
In the vertical direction, the domain extends from $\theta_{min} = 83^{\circ}$ to $\theta_{max} = 90^{\circ}$ (half disk), corresponding to 5 disk scale heights H in the initial setup, and in the $\phi$ direction from $\phi_{min} = 0^{\circ}$ to $\phi_{max} = 360^{\circ}$ (full disk). In the three coordinate system's directions $(r,\theta, \phi),$ we used a $600\times64\times1024 $ grid resolution. The grid cells are spaced logarithmically in the radial, and linearly in the other directions. The main parameters of the simulations are summarized in Table~\ref{tabel:model_param}.

The initial density profile created by force equilibrium is given by
\begin{equation}
    \rho(R,Z)=\rho_0\, \left(\frac{R}{R_\mathrm{p}}\right)^{p}\, \exp{\left[\frac{G M_\mathrm{s}}{c_\mathrm{s}^2}\left(\frac{1}{r}-\frac{1}{R}\right)\right]}\,,
    \label{eq:surfprof}
\end{equation}
where $R = r\sin{\theta}$ describes the cylindrical radius, $\rho_0$ is the gas mid-plane density at the planet location $R = R_\mathrm{p} = 1$, $p=-1.5$ the density exponent, and $c_\mathrm{s}$ is the isothermal sound speed. The disk's initial aspect ratio was set to $h=H/R= 0.05$, which corresponds to a temperature profile of
\begin{equation}
  T(R) = T_0\, \left(\frac{R}{R_\mathrm{p}}\right)^{q},
  \label{eq:tempprof}
\end{equation}
with $q=-1$ and $T_0$ = \SI{121}{K}.

The gas moves with an azimuthal velocity given by the Keplerian speed around a \SI{1}{M_\odot} star, corrected by the pressure support \citep{nelson2013}:
\begin{equation}
  \Omega(R,Z) = \Omega_\mathrm{K} \left[ (p +q) h^2 + (1+q) - \frac{qR}{\sqrt{R^2 + Z^2}} \right]^{\frac{1}{2}} \,,
  \label{eq:omega}
\end{equation}
where $\Omega_\mathrm{K}$ is the Keplerian orbital frequency.
At the inner and outer boundary, we adopted reflective conditions, and we damped the density as well as the radial and vertical velocity to the initial values at the timescale of a fraction of a local orbit in order to prevent reflection of the spiral wave caused by the planet-disk interaction onto the boundary. The damping was applied in the intervals $[0.4,0.5]$ $r_\mathrm{p}$ and $[2.3,2.5]$ $r_\mathrm{p}$.
For the vertical boundaries, a mirror condition was implemented at the disk midplane, while an open boundary was applied in the disk atmosphere, and in the azimuthal direction a periodic condition was applied.
We adopted a constant of $\alpha = 5 \cdot 10^{-4}$, which is consistent with values generated by hydrodynamical turbulences \citep[see e.g.,][]{stoll2017}.

\begin{table}
 \begin{center}
 \caption{Model parameters \label{tabel:model_param}}
 \centering
 \begin{tabular}{|c c|}
 \hline
 Parameter & Value \\ [0.3ex]
 \hline
 \textbf{Grid} &  \\ [0.3ex]
 \hline
 Radial range [au] & $2.08$ -- $13$ \\ [0.3ex]

 Vertical range [H] & $5$ \\ [0.3ex]

 Azimuthal range [rad] & $2 \pi$ \\ [0.3ex]

 Radial resolution & $600$ \\ [0.3ex]

 Polar resolution & $64$ \\ [0.3ex]

 Azimuthal resolution & $1024$ \\ [0.3ex]
 \hline
 \textbf{Gas} &  \\ [0.3ex]
 \hline
 Density index $p$ & $-1.5$ \\ [0.3ex]

 Temperature index $q$ & $-1.0$ \\ [0.3ex]

 Eqn. of state & Ideal, Isothermal \\ [0.3ex]

 $H/R$ & $0.05$, $0.02$ \\ [0.3ex]

 $\alpha$ & $5\cdot10^{-4}$ \\ [0.3ex]

 $\gamma$ & $1.4$, $1$ \\ [0.3ex]

 $\mu$ & $2.35$ \\ [0.3ex]
 \hline
 \textbf{Planet} &  \\ [0.3ex]
 \hline
 Planet mass [$M_{\oplus}$] & $5$, $10$, $100$ \\ [0.3ex]
 \hline
 \textbf{Dust} &  \\ [0.3ex]
 \hline
 Particle size [cm] & $0.01, 0.1, 1, 10, 30$ \\ [0.3ex]

    & $100, 300, 10^{3}, 10^{4},10^{5} $ \\ [0.3ex]

 Total number & $10^{6}$ \\ [0.3ex]
 \hline
 \end{tabular}
\end{center}
\end{table}

We employed the radiation hydrodynamics module developed by \cite{kolb2013radiation} for the {\sc PLUTO} code \citep{mignone2007pluto}. The solver is based on the FLD approximation in the two-temperature approach. The equations were solved in the comoving frame in the frequency-independent (gray) approximation.

The motion of the gas is described by the Navier-Stokes equations (eqs.~\ref{eq:gas_evol_1}, \ref{eq:gas_evol_2}, \ref{eq:gas_evol_3}) that are coupled with radiation transport (eq.~\ref{eq:evol_rad})
\begin{align}
&\frac{\partial}{\partial t}\rho + \nabla \cdot (\rho \bm{\upsilon}) = 0 \label{eq:gas_evol_1}\,,\\
&\frac{\partial}{\partial t}\rho \bm{\upsilon}+\nabla \cdot (\rho \bm{\upsilon} \otimes \bm{\upsilon} - \bm{\sigma}) + \nabla p = - \rho \nabla \Phi \label{eq:gas_evol_2}\,,\\
&\frac{\partial}{\partial t} e +\nabla \cdot [(e+P)\bm{\upsilon}] = - \rho \bm{\upsilon}\cdot \nabla \Phi + (\bm{\sigma}\cdot\nabla)\bm{\upsilon} - \kappa_{p} \rho c (a_{R}T^{4}-E) \label{eq:gas_evol_3}\,,\\
&\frac{\partial}{\partial t} E +\nabla \cdot \mathbf{F} = \kappa_{p} \rho c (a_{R}T^{4}-E) \label{eq:evol_rad}\,,
\end{align}
where the first three equations describe the evolution of the gas motion where $\rho$ is the gas density, $P$ corresponds to the thermal pressure, $\bm{\upsilon}$ is the velocity, $e=\rho \epsilon + 1/2 \rho \upsilon^{2}$ the total energy density (i.e., the sum of internal and kinetic energy) of the gas without radiation, $\epsilon = C_{V} T$ is the specific internal energy in which $C_{V}$ is the specific heat capacity (assumed constant here), $\bm{\sigma}$ is the viscous stress tensor, and $\Phi$ is the gravitational potential. This system of equations is closed by the ideal gas equation of state
\begin{equation}
     P = (\gamma - 1) \rho \epsilon = \rho \frac{k_{B}T}{\mu m_{H}}\,,
     \label{eq:ideal_gas}
\end{equation}
where $\gamma$ is the ratio of specific heats, T represents the gas temperature, $k_{B}$ is the Boltzmann constant, $\mu = 2.35$ the mean molecular weight, and $m_{H}$ is the hydrogen mass.

The evolution of the radiation energy density E is given by equation \ref{eq:evol_rad}, where $\mathbf{F}$ denotes the radiative flux, which was computed in the FLD approximation as in \citet{kolb2013radiation}, $\kappa_{P}$ is the Planck mean opacity, $c$ corresponds to the speed of light, and $a_{R}$ is the radiative constant.
This implementation does not include the advective transport terms for the radiation energy and radiative pressure work in equations \ref{eq:gas_evol_3} and \ref{eq:evol_rad}, since these terms are of minor importance for low-temperature disks.
For the computational boundaries, we adopted, for the radiative part, reflective conditions in the radial direction, a symmetric condition at the disk midplane, and a fixed temperature of \SI{5}{K} in the upper boundary. Period boundaries were applied in the azimuthal direction.

\subsection{Dust component}\label{subsec:dust_component}

The solid fraction of the disk is modeled with $10^6$ Lagrangian particles divided into ten size bins as reported in Table~\ref{tabel:model_param}. This approach has the great advantage of modeling a broad range of dynamical behaviors self-consistently, using the same modeled particles. The trade-off is that in the regions of low density, the resolution of the dust population is lower. However, for our study, this is not a problem since we are mainly interested in the dynamical evolution of dust particles; thus we do not take into account collisions between particles or the back-reaction of the dust onto the gas. We study particles with sizes $s$ from \SI{0.1}{mm} up to \SI{1}{km} and internal density $\rho_\mathrm{d} =$ \SI{1}{g\, cm^{-3}}. The particle sizes were chosen to cover a wide range of different dynamical behaviors. The initial surface density profile of the dust particles is
\begin{equation}
\Sigma_\mathrm{d}(r) \propto R^{-1}.
\end{equation}
This particle distribution leads to an equal number of particles in each radial ring as the grid is spaced logarithmically in the radial direction. The resulting dust profile is steeper than the gas profile, and it allows for a better sampling of the dust dynamics in the vicinity of the planet location. The dust particles were initially placed with a vertical distribution given by the local disk scale height and the dust diffusion coefficient \citep[see e.g.,][]{youdin2007}.
The evolution of dust particles is given by the gravitational interaction with the planet and central star, turbulent kicks from the gas that resemble a realistic turbulent behavior, and the drag force from the interaction with the gaseous disk
\begin{equation}
    \mathbf{F}_\mathrm{drag} = -\frac{m_\mathrm{d}}{t_\mathrm{s}}
    \mathbf{v}_\mathrm{r} \,,
\end{equation}
where $m_\mathrm{d}$ is the dust particle mass, and $t_\mathrm{s}$ is the stopping time which represents the timescale on which the embedded dust particle approaches the gas velocity and, for well-coupled particles, it is given by
\begin{equation}
    \label{eq:stop}
    t_\mathrm{s}=\frac{s\rho_\mathrm{d}}{\rho \bar{v}_\mathrm{th}} \,,
\end{equation}
or, in its dimensionless form (hereafter, Stokes number), as
\begin{equation}
\label{eq:stokes}
    \tau_\mathrm{s}=t_\mathrm{s}\Omega_\mathrm{K}(\mathbf{r}) \,,
\end{equation}
which describes the effect of a drag force acting on a particle independently of its location within the disk.
Here, $\bar{v}_\mathrm{th}$ represents the mean gas thermal velocity.
Dust particles were introduced at the beginning of the full 3D simulation in thermal equilibrium, and they evolve with two different integrators depending on their $\tau_\mathrm{s}$ \citep[see][ for further details]{picogna2018particle}. We do not consider the effect of the disk self-gravity on the particle evolution. Particles that leave the computational domain at the inner boundary re-enter at the outer boundary. This solution allowed us to keep a constant number of particles, which is beneficial from a numerical point of view, and it does not affect the load balance between the computational cores. Adding particles at a constant rate from the outer boundary would only make sense if we were considering the evolution of small planets for longer times. As shown in Fig.~\ref{fig:HistTime}, the disk close to the planet location does not run out of particles within the simulated time even for the fast evolving pebbles. Accreted particles are flagged but were otherwise kept in the simulations (see Sec.~\ref{sec:results} for further details).

\subsection{Planets}\label{subsec:planets}

We embed a planet, with a mass in the range of $[5, 10, 100]$ $M_\oplus$, orbiting a solar mass star on a circular orbit with semi-major axes $R_p = 1$ (\SI{5.2}{au}) in code units. The planet does not migrate and its mass is kept fixed. Its gravitational potential is smoothed with a cubic expansion inside its Hill sphere
\begin{equation}\label{eq:hill}
r_\mathrm{H} = r  \left(\frac{M_{p}}{3M_{\star}}\right)^{1/3}\,
\end{equation}
\citep{klahr20063d}.
In order to obtain the initial conditions, we evolved the disk for $\sim200$ orbits, with lower resolution in the azimuthal direction, until its scale height was not changing considerably. This allowed us to start our simulation with the disk in thermal equilibrium. Subsequently we increased the resolution to the one reported in Tab.~\ref{tabel:model_param}, and added the dust component.
After the dust component has been evolved for six orbits in the computational domain, the planetary mass slowly increases over an additional $20$ orbits to allow for a smooth initial phase.
Each simulation was evolved for $100$ orbital periods of the planet when the disk structure is close to a stationary state and further changes are small, so we do not expect them to affect our results.

\section{Results}\label{sec:results}

\subsection{Gas evolution}
\begin{figure}
    \includegraphics[width=\linewidth]{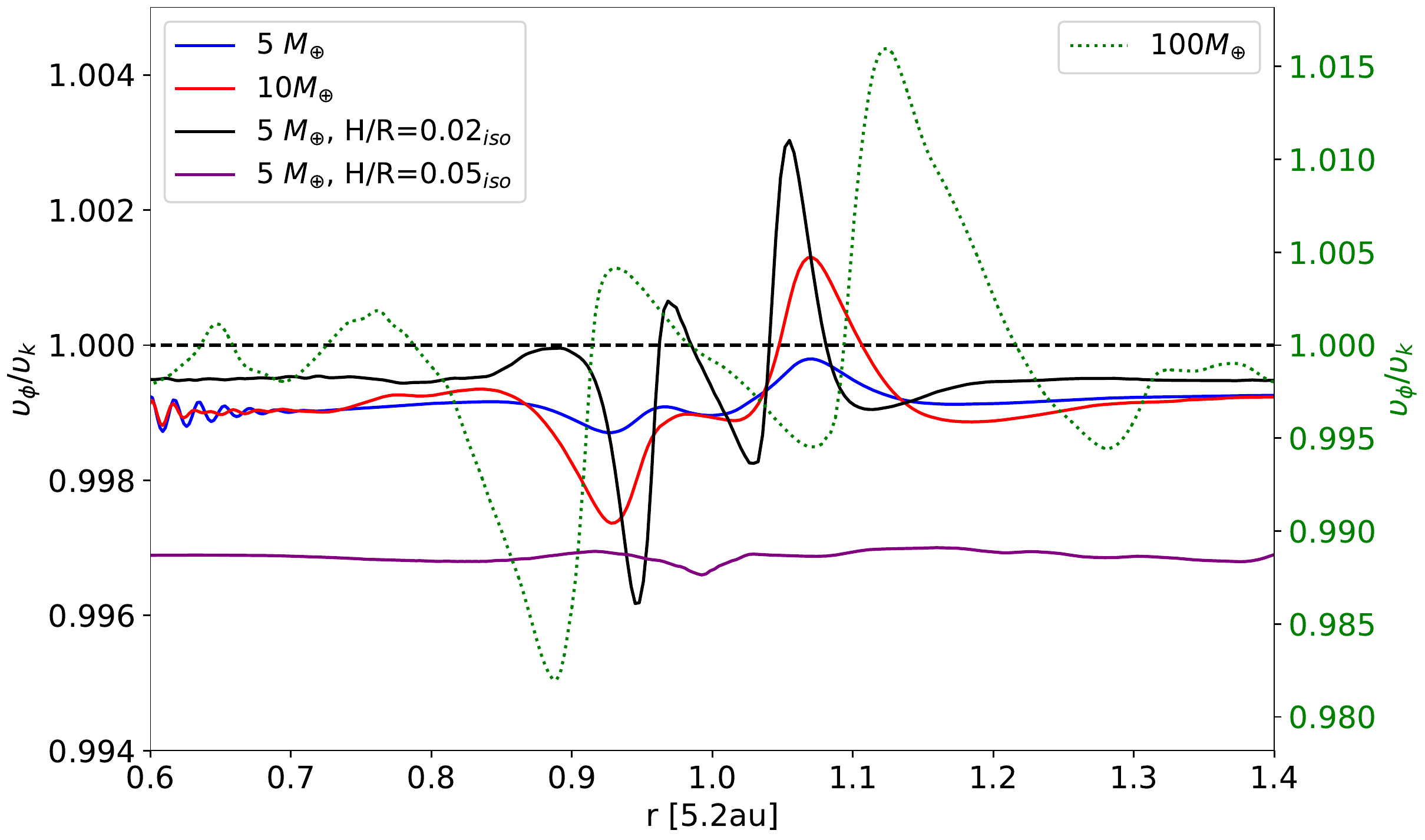}
    \caption[utheta-uk]{Azimuthal gas velocity in units of the Keplerian speed as a function of radius for the different planetary masses and models. When the gas speed becomes super-Keplerian outside the planet location, the dust-filtration process occurs and the pebble isolation mass is reached. The purple line shows the corresponding simulation from \cite{picogna2018particle} at the same time span. The \SI{100}{M_\oplus} planet corresponds to the right axis and it is shown with a green dotted line.}
    \label{fig:utheta-uk}
\end{figure}

The presence of a planet significantly alters the gas structure close to its location even if it is not massive enough to open up a gap \citep[see e.g.,][]{lin1993,lambrechts2014separating}.
In Fig.~\ref{fig:utheta-uk} we show the azimuthal gas velocity in units of the Keplerian speed as a function of radius. In this case, we compared our results with those of \citet{picogna2018particle} for a viscous locally isothermal disk with a disk scale height of $H/R=0.05$ and with an isothermal case with $H/R=0.02$. This helps us to understand the influence of radiative cooling.
The first thing to notice is that changing the disk aspect ratio changes the pressure support of the gas as well, so that the disk becomes more Keplerian on average, going from $H/R=0.05$ (purple line) to $H/R=0.02$ (black line).
This plays an important role in the dust evolution because whenever the gas becomes super-Keplerian, the head-wind felt by the dust particles is reduced, meaning that the particles slow their inward migration down  or stop it altogether.
Thus, for a colder (thinner) disk, smaller planets can halt the dust evolution more easily.
As seen from the parameter study probed, a \SI{5}{M_\oplus} planet can create a super-Keplerian flow outside its location for an $H/R=0.02$, while the flow remains sub-Keplerian everywhere for the same planet in a hotter disk ($H/R=0.05$).

\begin{figure}
    \includegraphics[width=\linewidth]{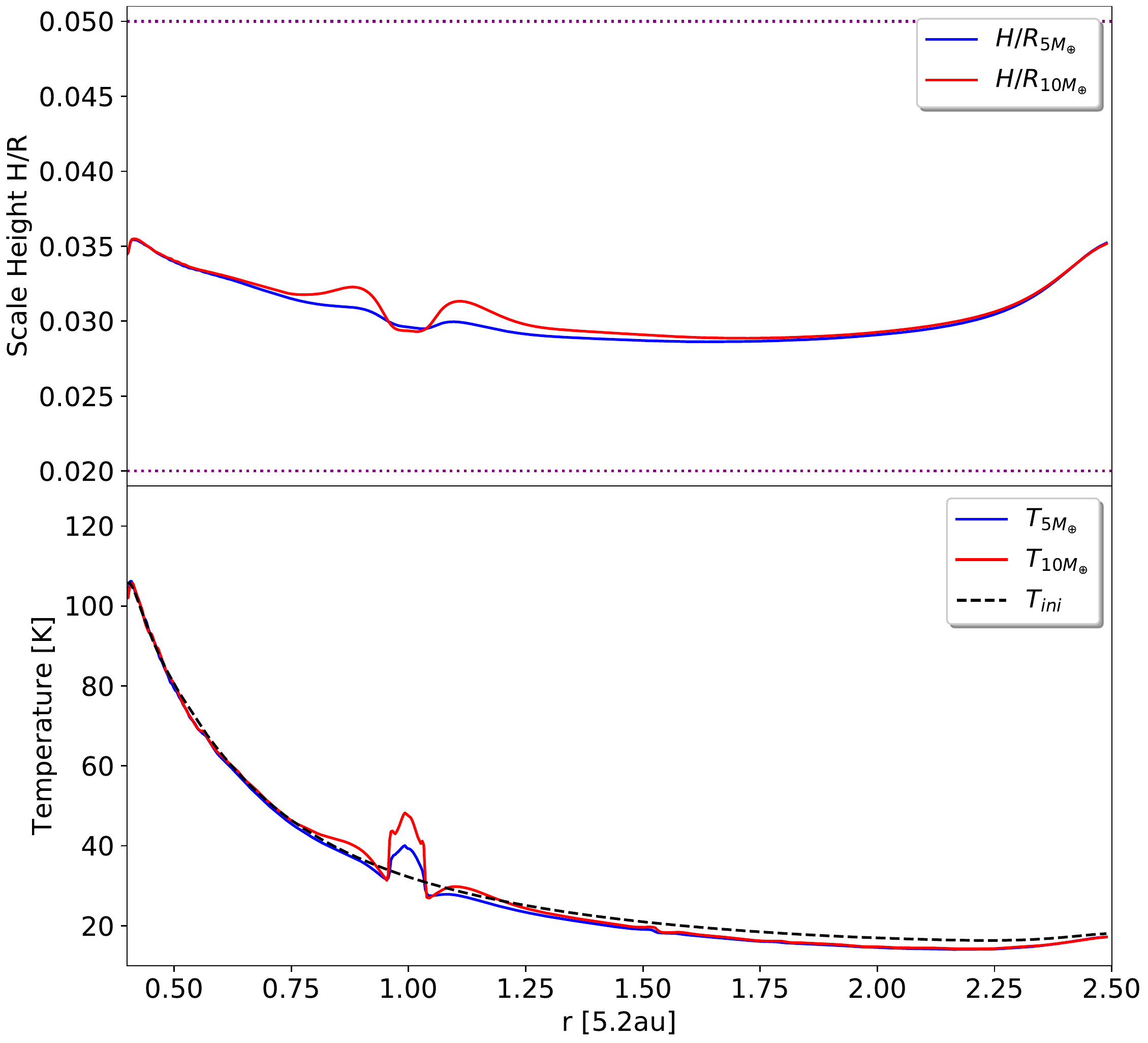}
    \caption[scale height]{\textit{Upper panel}: Disk scale height ($H/R$, solid lines) as a function of radius for the different planetary masses, $\SI{5}{M_\oplus}$ (blue) and $\SI{10}{M_\oplus}$ (red), for the models with radiative transfer. The purple dotted lines at $0.05$ and $0.02$  represent the $H/R$ of the isothermal simulations performed, respectively.
    
    \textit{Lower panel}: Midplane gas temperature (T, solid lines) for the corresponding models as on the top panel. The black dashed line represents the initial temperature.}
    \label{fig:scale_height}
\end{figure}

\begin{figure}
    \includegraphics[width=\linewidth]{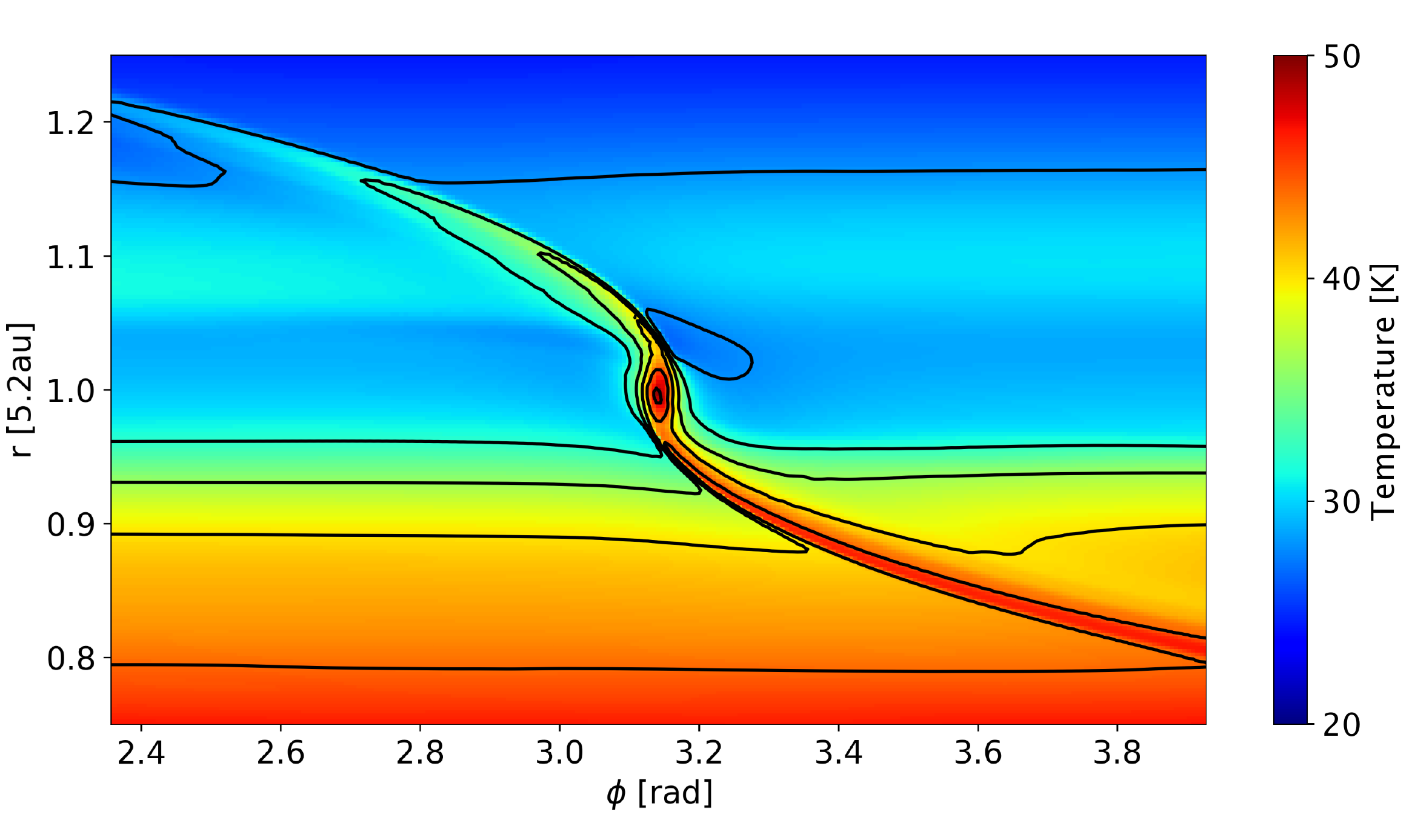}
    \caption[tem2d]{Gas temperature map at the disk mid-plane for the $\SI{10}{M_\oplus}$ planet. The temperature close to the planet location increases significantly.}
    \label{fig:temp_2d}
\end{figure}

The radiative case lies in the middle of the two cases presented above even though the initial aspect ratio of the disk was $0.05$ in this case and the \SI{5}{M_\oplus} planet almost pushes the gas to the Keplerian speed (blue line).\ This points to the fact that the disk was able to effectively cool and adapt to a new equilibrium aspect ratio.
This effect is highlighted in Fig.~\ref{fig:scale_height} where the two isothermal cases are plotted with purple dotted lines for reference. The radiative case settles to a disk scale height that is close to $0.03$ at the planet location, with small perturbances in the planet vicinity depending on its mass.
The features in the disk scale height can be directly related to the temperature structure (shown in Fig.~\ref{fig:scale_height}, lower panel).
There, we see that the region close to the location of the planet is significantly hotter than the surrounding disk. We show this effect in more detail in Fig.~\ref{fig:temp_2d} where we plotted the temperature map at the disk mid-plane for the $\SI{10}{M_\oplus}$ planet. For a planet at a larger separation, this effect could be even more pronounced and indirect evidence for the planet \citep[see e.g.,][]{Tsukagoshi2019}.

\subsection{Dust distribution}
\begin{figure}
    \centering
    \includegraphics[width=\linewidth]{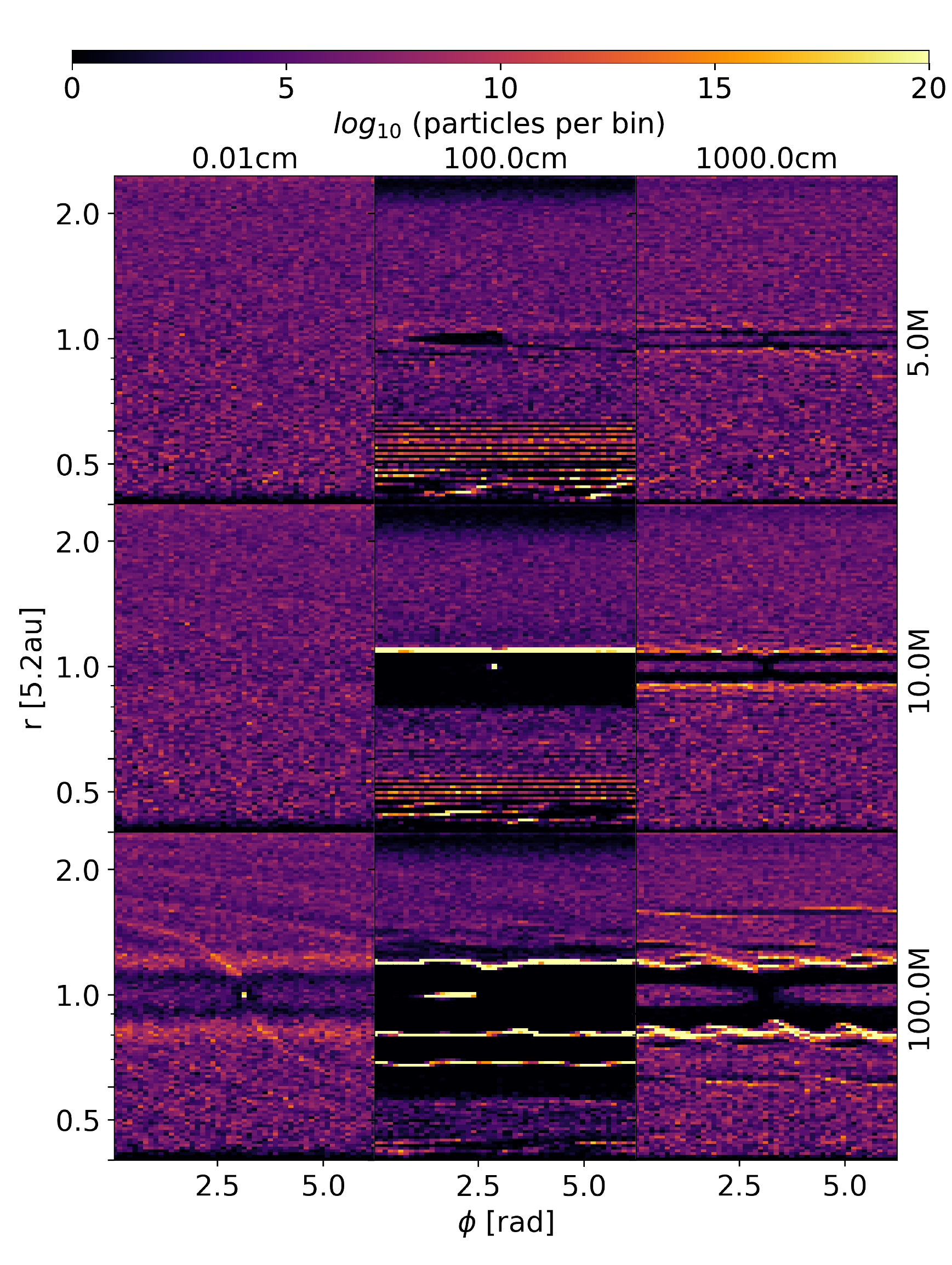}
    \caption[Spatial distribution]{Spatial distribution of the dust particles after 100 planetary orbits for the different planetary masses and for three representative particle sizes. The Stokes numbers at the planet location from left to right are, $0.08$, $1.23$, and $67.2,$ respectively.}
    \label{fig:HistDist}
\end{figure}

The different particle sizes modeled encompass a wide range of stopping times. In Figure \ref{fig:HistDist} we show the dust surface number density after $100$ planetary orbits for three representative sizes and the three planetary masses.
In particular, the middle column shows the distribution of pebble-sized particles, which are defined here as particles with a Stokes number of order unity and correspond to meter-sized objects in our setup.

The \SI{5}{M_\oplus} planet (first row) is not massive enough to open up a gap in the gas and particle disk as seen for the small (coupled) dust in the left column. The only exception is for the planetesimal-sized objects (right column), which do not feel a strong gas drag, and the planetary core can gravitationally perturb their orbits, depleting the co-orbital region \citep{dipierro2017}.
In the middle row, on the other hand, the \SI{10}{M_\oplus} planet has stopped the inflow of pebble-sized particles (middle column), creating an overdensity at the location of the outer pressure bump and a depleted inner disk \citep{paardekooper2006,rice2006,pinilla2012}.
The \SI{100}{M_\oplus} planet is able to carve a deep gap in  both the gas and the small particles (last row, first column), where the spiral arms launched by the planet are visible as well. The gap edges also become Rossby wave unstable, generating vortices \citep{lovelace1999}, as is visible in the larger particle distribution. In the pebble-sized objects (last row, middle column), the distribution of solids shows two concentric rings in the inner side of the planet caused by the perturbance in the gas Keplerian speed by the massive planets, which generates two confined regions where the gas becomes super-Keplerian (see Fig.~\ref{fig:utheta-uk}).

\subsection{Dust evolution}
\begin{figure}
    \centering
    \includegraphics[width=\linewidth]{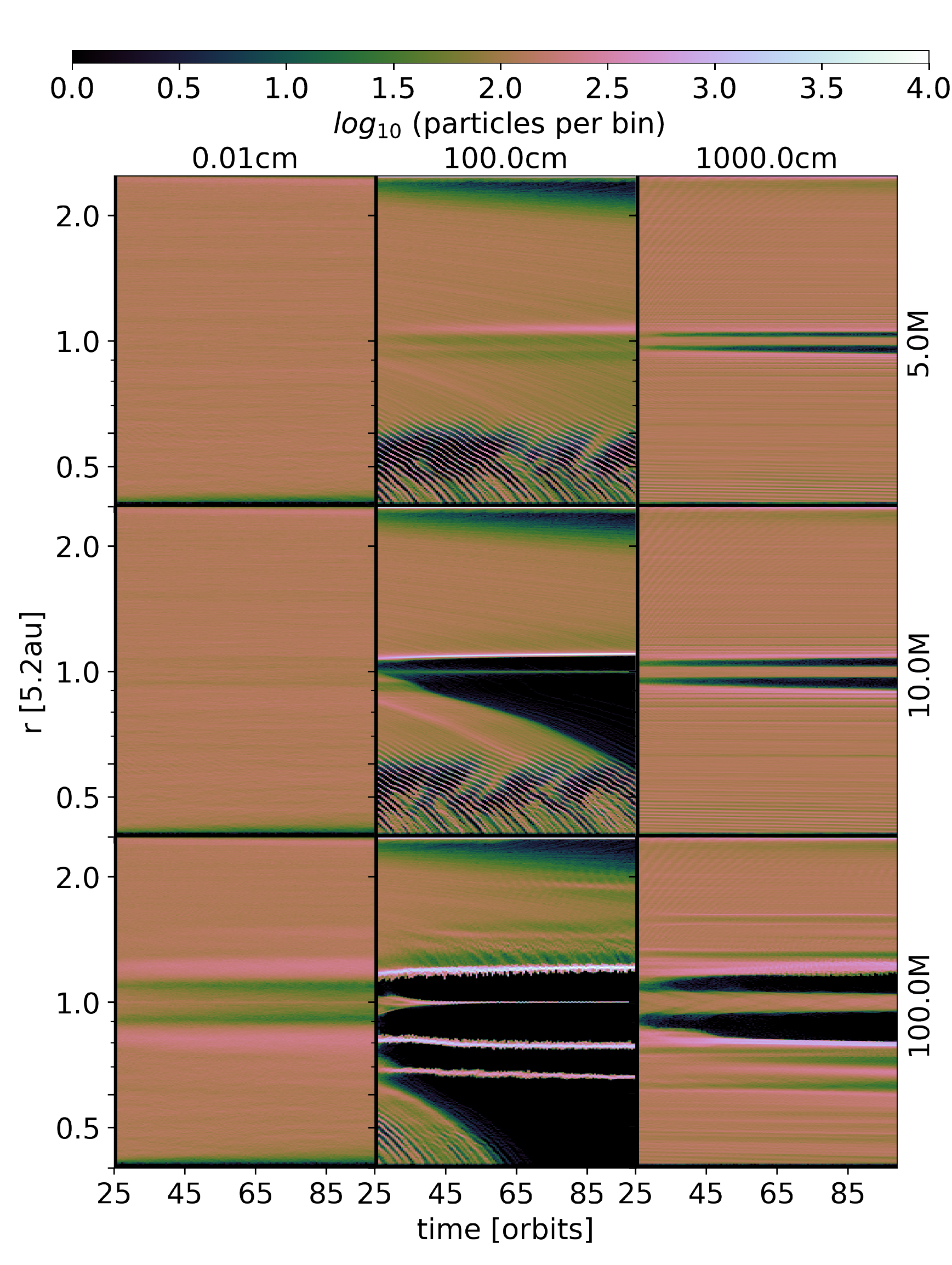}
    \caption[Radial distribution]{Radial distribution (averaged over azimuth) of the dust particles as a function of time for the different planetary masses and for three representative particle sizes. The \SI{10}{M_\oplus} planet is able to stop the pebbles efficiently, thus creating a transition disk, suggesting that the pebble isolation mass can be less than \SI{10}{M_\oplus}. The Stokes numbers at the planet location from left to right are $0.08$, $1.23$, and $67.2$  after $100$ planetary orbits, respectively.}
    \label{fig:HistTime}
\end{figure}

In Figure~\ref{fig:HistTime} we show the temporal evolution of the radial distribution for the same dust particles presented earlier as a function of planetary mass. The plot shows that a quasi-equilibrium state has been obtained for the dust distribution at the end of the simulated period.
The \SI{5}{M_\oplus} planet (first row) is not able to stop the radial inflow of the well-coupled dust particles (left column), while a small overdensity is building up outside its location for the pebble-sized particles (middle column). On the other hand, a \SI{10}{M_\oplus} planet is able to stop the flux of pebble-sized objects efficiently (second row, middle column), thus creating a transition disk where the inner disk is depleted of dust as they are not able to cross the pressure bump created by the planet. The mass at which a planet is able to effectively stop the influx of pebble-sized objects is called pebble isolation mass \citep[see e.g.,][]{morbidelli2012}.
This effect becomes even stronger for the \SI{100}{M_\oplus} planet (last row) where already after $60$ planetary orbits, the pebble particles (middle column) inside the planet location are only present in two confined regions (as described earlier). In looking at the right column, one can see the strong mass dependence of the gap width for planetesimal-sized objects as well \citep{ayliffe2012,weber2018}.

\subsection{Solid accretion}
The solid accretion rate onto the planetary cores is computed in a post-processing phase using two different criteria, depending on the ratio between their Stokes number and the time they need to cross the planet's Hill sphere \citep[see][ for a detailed description]{picogna2018particle}.
A sample of particles for each size was chosen outside the planet location and followed for $50$ planetary orbits, after which we checked the fraction of accreted and nonaccreted particles \citep[see Fig. 10 of ][]{picogna2018particle}.
Mass is not added to the planet, and even if accreted particles can remain close to the planet or reach a high speed, they do not influence the simulation since the interaction between dust particles and back reaction onto the gas are neglected.
We verified that the total solid accretion rate reached a quasi-equilibrium state after $100$ planetary orbits when the simulation was stopped.

We can then compute an effective accretion efficiency, $P_\mathrm{eff}$, that is, the number of accreted particles onto the planet divided by the number of particles that would otherwise drift across the location of the planet in an unperturbed disk \citep{ormel2010}
\begin{equation}
    P_\mathrm{eff} = \frac{\dot{M}_\mathrm{acc}}{\dot{M}_\mathrm{drift}},
\end{equation}
where $\dot{M}_\mathrm{acc}$ is the measured accretion rate through the criteria explained above, while $\dot{M}_\mathrm{drift}$ is the particle drift through the disk $\dot{M}_\mathrm{drift} = 2\pi r \Sigma_\mathrm{p} v_\mathrm{drift}$, where $\Sigma_\mathrm{p}$ is the particle surface density and $v_\mathrm{drift}$ is the unperturbed dust radial drift speed \citep[see e.g.,][]{nakagawa1986}.

\begin{figure*}
    \centering
    \includegraphics[width=\linewidth]{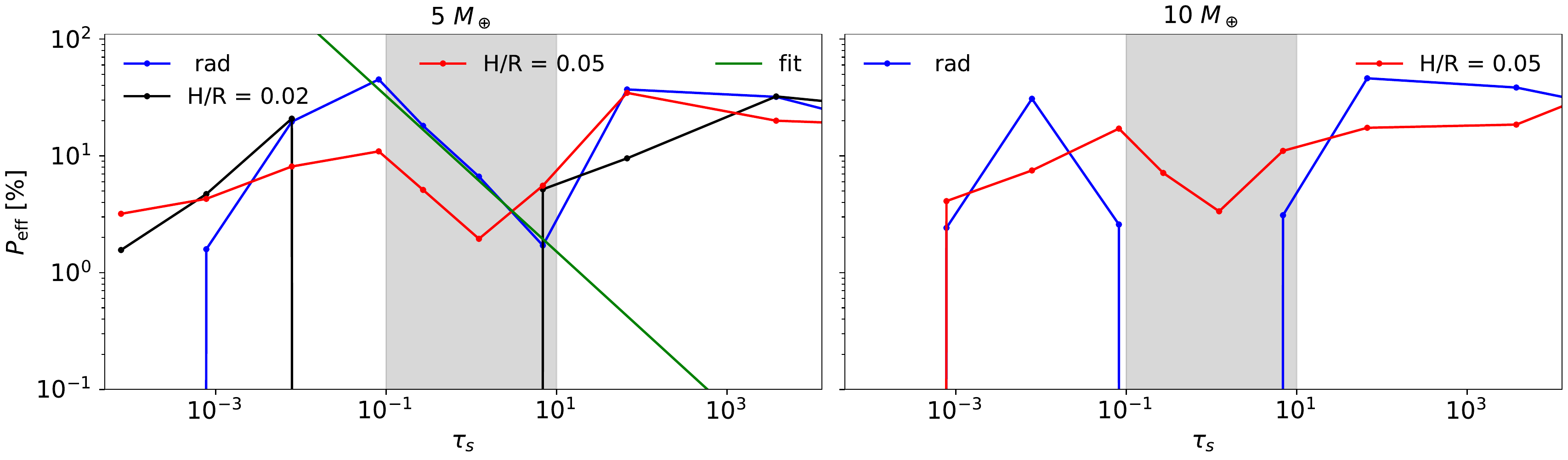}
    \caption[Efficiency]{Efficiency of accreted particles as a function of Stokes number for different planetary masses. The blue line (rad) refers to the new simulations performed with radiative transfer, the black line corresponds to the isothermal case with $H/R=0.02$, and the red line ($H/R=0.05$) represents the isothermal case from \cite{picogna2018particle}. The fit from eq.~\ref{eq:peff} is overplotted with a green line. We shaded the intermediate $\tau_{s}$ values in gray for which the method adopted to calculate the efficiency holds (see text).}
    \label{fig:efficiency}
\end{figure*}

In Figure~\ref{fig:efficiency}, we plotted the accretion efficiency for two different planetary masses, \SI{5}{M_\oplus} and \SI{10}{M_\oplus}, and we excluded the \SI{100}{M_\oplus} planet because it is capable of carving a gap very quickly, thus strongly depleting the solid accretion for all the studied particle sizes. Again, we compare the radiative case (blue) with two locally isothermal ones with different aspect ratios (red \& black). As already pointed out by \citet{picogna2018particle}, we can only trust the middle part of this plot (in the Stokes number range from $0.1$ to $10$) because for larger or smaller particles, the planet moves faster than the dust and the accretion efficiency cannot be computed correctly with our method, since the sample of particles outside the planet location chosen for the efficiency calculation is not able to cross its location. Therefore, we highlight the relevant region with a gray color.

\begin{figure}
    \includegraphics[width=\linewidth]{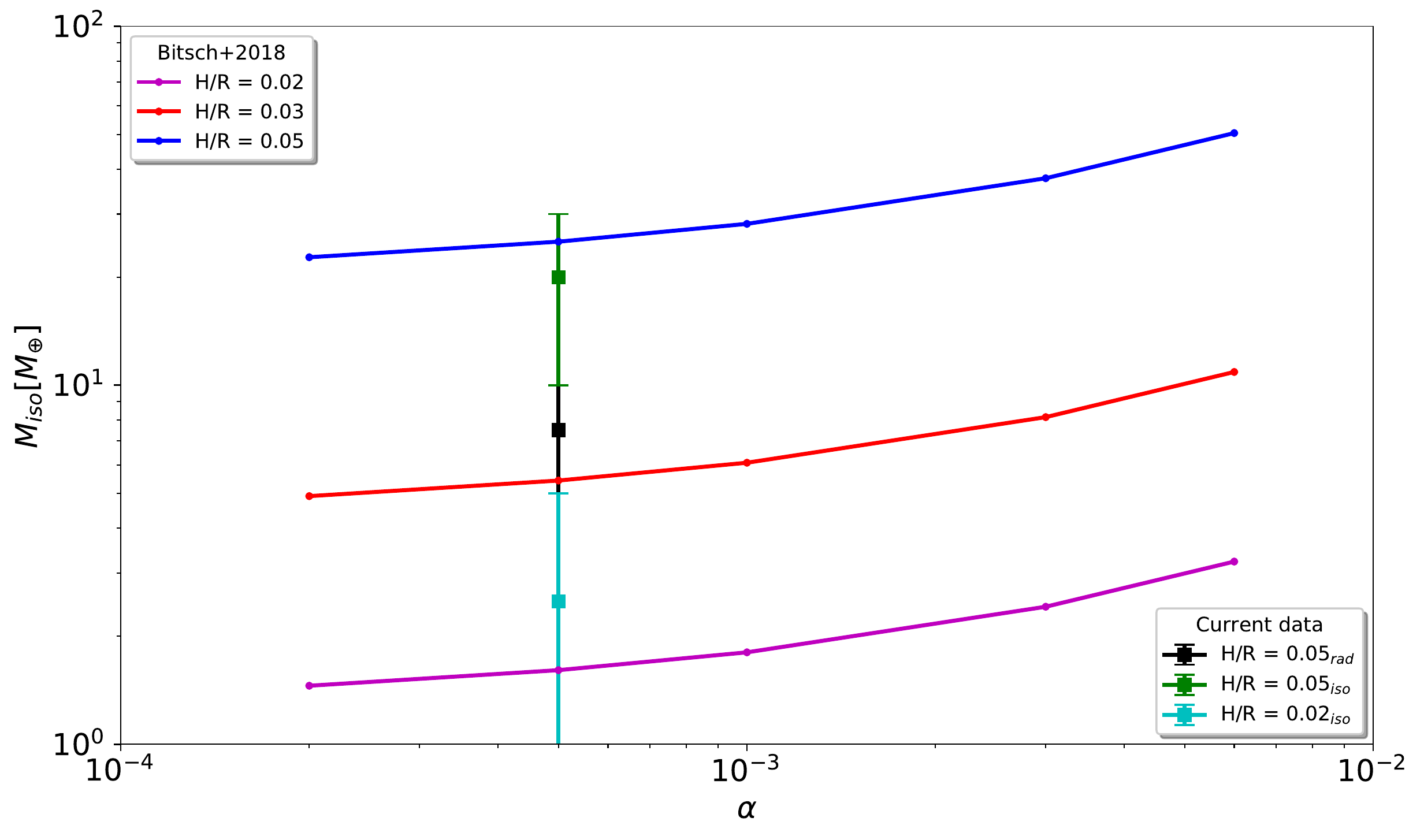}
    \caption[p isolation]{Pebble-isolation mass as a function of $\alpha$ and for different aspect ratios $H/R$. The lines show the pebble isolation mass for different $\alpha$ values and they follow the fit from \citet{bitsch2018pebble}. Squares and error bars correspond to the data from \citet{picogna2018particle} and the new models with radiative transfer.}
    \label{fig:p isolation}
\end{figure}

For the \SI{5}{M_\oplus} planet (Fig.~\ref{fig:efficiency}, left panel), the accretion efficiency in the radiative case (blue line) has a similar steepness but it has increased with respect to the locally isothermal case with $H/R=0.05$ (red line), by a factor of three to four; whereas, for the case with a lower aspect ratio, the "pebble isolation mass" has already been reached, effectively stopping the inflow of pebble-sized particles.
The difference becomes striking for the \SI{10}{M_\oplus} planet (Fig.~\ref{fig:efficiency}, right panel). The pebble isolation mass is reached for the radiative case (blue line), and the overall accretion efficiency is reduced. While in the isothermal case with a higher aspect ratio (red line), the planet is still accreting pebble-sized objects efficiently.
The reason for this difference can be attributed to the change in the disk aspect ratio at equilibrium for the radiative disk compared to the locally isothermal ones (see Fig.~\ref{fig:scale_height}). This effect has been predicted by \citet{bitsch2018pebble}, who found that the pebble isolation mass scales strongly with the disk aspect ratio for a locally isothermal disk
\begin{equation}
    M_\mathrm{iso} \propto \left(\frac{H/R}{0.05}\right)^3.
\end{equation}
In Figure~\ref{fig:p isolation} we compare the pebble isolation mass found for our simulations (with error bars given by the bin size in planetary mass used) with the analytical prescription from \citet{bitsch2018pebble}, and we find good agreement when taking the reduced disk scale height at equilibrium for the radiative case into account.

We fit the accretion efficiency with an analytical prescription in the Stokes number range considered, (see green line in Fig.~\ref{fig:efficiency}) as
\begin{equation}\label{eq:peff}
P_\mathrm{eff} = \frac{3}{\pi\eta}\left(\frac{\tau_s}{0.1}\right)^{-2/3} \left(\frac{r_\mathrm{H}}{r}\right)^2\,
\end{equation}
where $r_\mathrm{H}$ is the Hill radius, and
\begin{equation}\label{eq:eta}
\eta = -\frac{1}{2}  \left(\frac{H}{r}\right)^2  \frac{\partial \ln P}{\partial \ln r} = 8.56\cdot 10^{-4} 
\end{equation}
is the value at $\SI{5.2}{au}$ in our simulation, just before the planet was included.

This relation is similar to the one obtained by \citet[][see their eq.~33]{lambrechts2014forming}, but it shows a steeper dependence on the Stokes number $\tau_s$. Their analytical derivation gives a correlation to the Stokes number as $P_\mathrm{eff} \propto {\tau_s}^{-1/3}$.
There are two main reasons to explain why our study yields a steeper relation. First of all, their relation only holds up to $\tau_s = 0.1$. This is because the ratio of the pebble accretion rate \citep[see eq.~28 of ][]{lambrechts2014forming}, given by $\dot{M_{c}} \propto \tau_{s}^{2/3}$, and the pebble flux through the planet $\dot{M_{f}} \propto \upsilon_{r} \propto \frac{\tau_{s}}{\tau_{s}^{2}+1} \simeq \tau_s$ gives this proportionality. However, this last relation breaks down for $\tau_{s} > 0.1$. In our case, the proportionality of $-2/3$ holds for all of the values of $\tau_{s}$ that are in agreement with our method ($\tau_{s}$ from 0.1 - 1.0, gray area in Fig. \ref{fig:efficiency}).
Furthermore, we are treating a full 3D disk with turbulent kicks on the dust particles, which is difficult to model using an analytic prescription, as in \citet{lambrechts2014forming}. In our 3D simulations, the small particles are lifted to a higher disk scale height compared to the larger particles that settle in the midplane. Hence, the accretion rate increases for smaller dust particles, and a steeper gradient of the efficiency as a function of the particle Stokes number develops.



\section{Conclusions}\label{sec:conclusions}

We can summarize the main results of this study with the following points.

\begin{itemize}
  \item The main effect of radiative transfer within the bulk of the disk is to produce a cooler and thinner disk (lower $H/R$, see also \citet{kley2009}). This allows the growing protoplanet to enhance its efficiency at halting the flux of pebbles and decrease the pebble isolation mass.
  Since this limit becomes lower than $10 M_{\oplus}$, which corresponds to the mass at which the core can start its runaway gas accretion, this result can simply explain the observed, relatively low gas giant planet frequency.
  \item We found a new relation for the pebble efficiency as a function of the particle Stokes number for a 3D disk (see eq.~\ref{eq:peff}), where the dependence on the Stokes number is steeper than the one obtained by the analytical derivation of \citet{lambrechts2014separating}.
  \item The planetary cores at the end of the disk lifetime would have had time to grow until this limit, which is consistent with the finding that the majority of exoplanets are in the mass range between Earth and Neptune.
  \item Decreasing the limit of planetary masses that are able to stop the inflow of pebbles from the outer to the inner disk, with respect to the planet location, can help to explain the population of transition disks observed around young stars (with high accretion rates), in a natural way.
\end{itemize}

The main limitations of this study are the lack of dust back-reaction, particle growth, and irradiation from the central star.
Dust coagulation and fragmentation can increase the fraction of solids crossing the planet location by grinding them down at the pressure bump and regrowing them in the inner disk \citep{Drazkowska_2019}. Along the same lines, dust back-reaction would modify the pressure gradient outside the planet location, reducing the planet efficiency in stopping the pebble flux \citep{kanagawa2018}. In this respect, our work, which does not take these two effects into account, underestimates the pebble isolation mass. However, \citet{Drazkowska_2019} found that the effect of back-reaction can be strongly reduced when taking a full coagulation model into account.
Stellar irradiation is expected to play an important role in modifying the disk aspect ratio and potentially the pebble isolation mass.
This effect is, nevertheless, only stronger in the outer, irradiation dominated, disk.
We looked in detail to understand the impact of stellar irradiation in Appendix~\ref{app:hr}, confirming that its contribution at the location studied is negligible. 
We did focus on a planet at \SI{5.2}{au} where viscous heating is still the main heating term \citep[see e.g.,][]{otherAZ}, and so our results can be safely generalized to a more complex model.

The fact that our conclusions do not hold for larger radii could imply that giant planets only form in the irradiation dominated region, and the signature of this effect can still be present in the bulk composition of the observed giant planets.
Finally, we modeled a laminar viscous disk. It has been shown that, even in magnetically inactive regions, hydrodynamical instabilities can develop \citep[see e.g.,][ for the vertical shear instability, VSI]{stoll2017}.
As seen by \citet{stoll2016}, VSI has a significant effect on reducing the effective viscosity and the disk scale height while increasing the dust vertical spreading. Its effect on solid accretion rates and isolation masses has been studied in the context of locally isothermal disks in \citet{picogna2018particle} who found a negligible difference. However, for radiative disks, the damping effect of VSI on the disk scale height can further strengthen our main conclusions.

\begin{acknowledgements}
We acknowledge funding by the DFG Research Unit 'Transition Disks' (FOR 2634/1, ER 685/8-1, KL 650/29-1), and by the European Research Council (ERC) under the European Union's Horizon 2020 research and innovation programme (grant agreement No 714769), and by the German Research Foundation cluster of excellence ORIGINS (EXC 2094, www.origins-cluster.de). The authors gratefully acknowledge the compute and data resources provided by the Leibniz Supercomputing Centre (www.lrz.de).
We thank Til Birnstiel and Joanna Dr{\c a}{\.z}kowska for proofreading the paper, and the anonymous referee for the constructive feedback.
\end{acknowledgements}

\bibliographystyle{aa}
\bibliography{references}

\appendix

\section{Importance of stellar irradiation}\label{app:hr}

\begin{figure}[htbp!]
    \includegraphics[width=\linewidth]{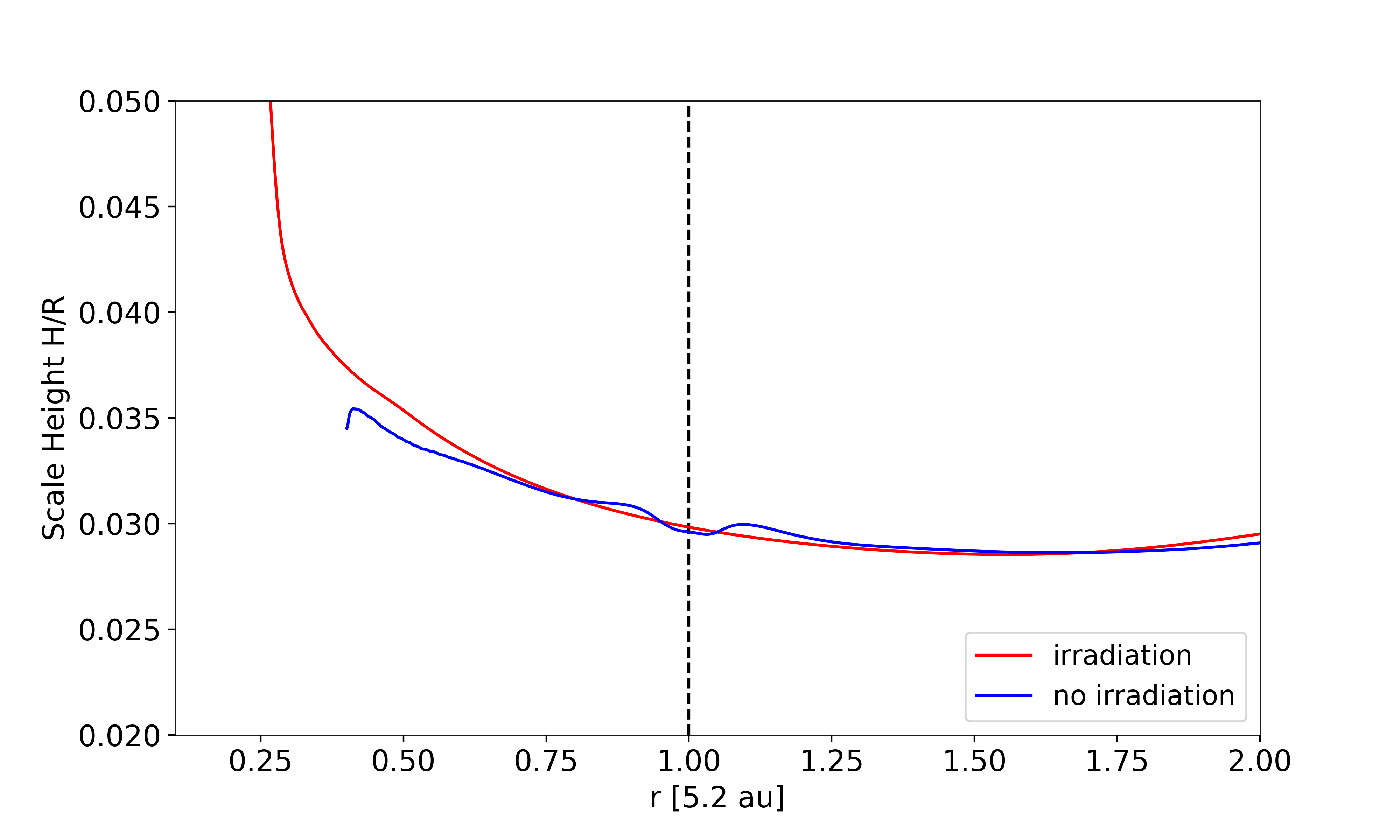}
    \caption[h/r]{Disk scale height for a stellar irradiated disk (red) compared with the disk scale height of the run with a \si{5}{$M_\oplus$} planet (blue).}
    \label{fig:scale_heigh_irr}
\end{figure}

\begin{figure}[htbp!]
    \includegraphics[width=\linewidth]{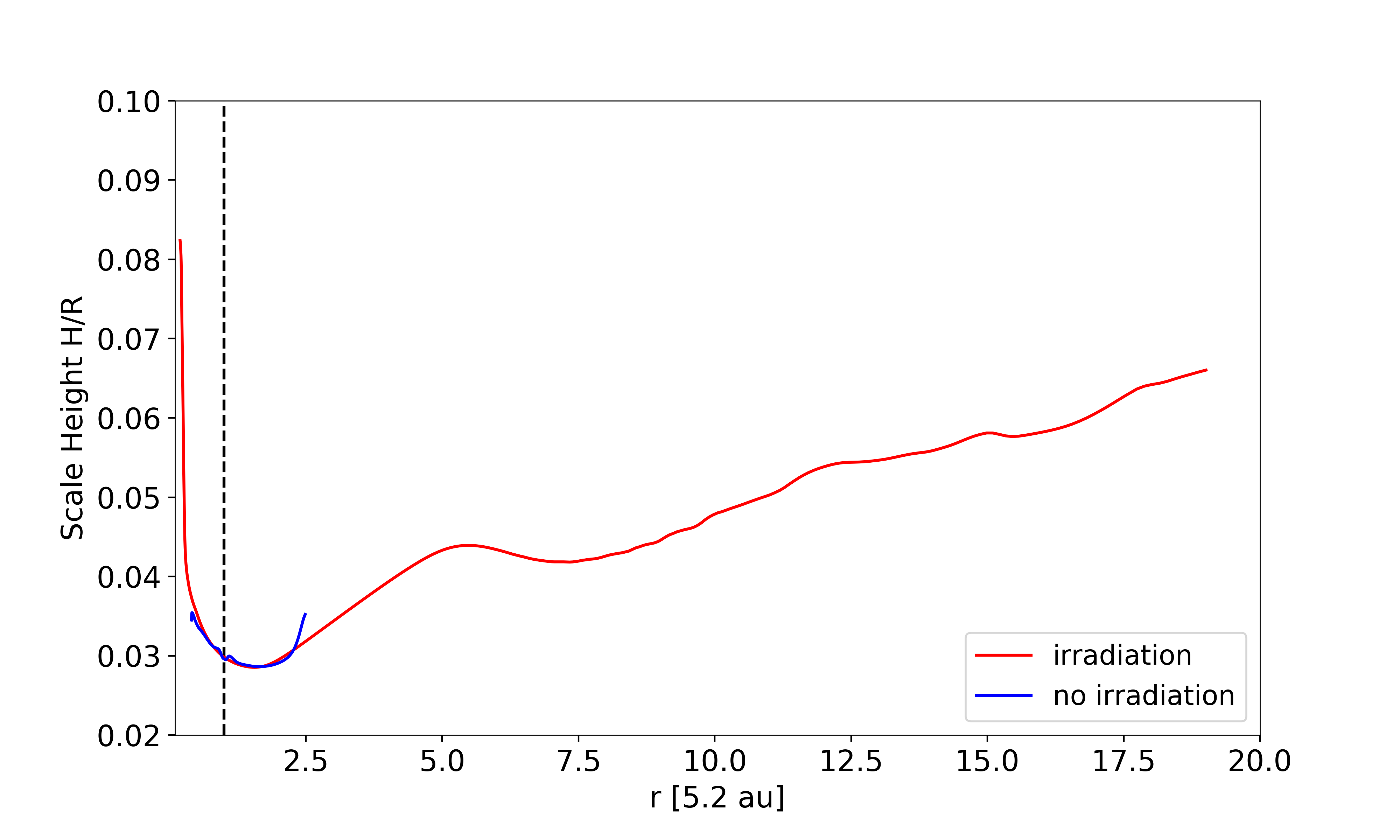}
    \caption[h/r ext]{Disk scale height for a stellar irradiated disk with a \si{100 } {au} outer radius (red) compared with the disk scale height of the run with a \si{5}{$M_\oplus$} planet (blue).}
    \label{fig:scale_height_ext}
\end{figure}

In this work, we ignore the contribution of the stellar irradiation to the disk temperature structure. Since one of our main results is the reduction of the pebble isolation mass due to the decreased disk aspect ratio, we tested our assumption. We performed an extra run (without planet and dust particles) including stellar irradiation (with $T_\star=5772$ K), until thermal equilibrium was reached at the planet location. As shown in Fig.~\ref{fig:scale_heigh_irr}, the disk scale height at the planet location is unaffected by the contribution of the stellar irradiation (red line) when comparing it to the run for the \si{5}{$M_\oplus$} planet (and no stellar irradiation, blue line). On the other hand, this effect plays a major role in an extended disk (see Fig.~\ref{fig:scale_height_ext}).

\end{document}